\documentclass[manuscript,nonacm]{acmart}

\usepackage{booktabs}

\usepackage{array}
\newcommand{\PreserveBackslash}[1]{\let\temp=\\#1\let\\=\temp}
\newcolumntype{C}[1]{>{\PreserveBackslash\centering}p{#1}}
\newcolumntype{R}[1]{>{\PreserveBackslash\raggedleft}p{#1}}
\newcolumntype{L}[1]{>{\PreserveBackslash\raggedright}p{#1}}

\usepackage{amsmath}
\usepackage{bm}
\usepackage{amsfonts}
\usepackage{amsthm}
\usepackage{multirow}
\usepackage{colortbl}
\usepackage{enumitem}
\usepackage{subcaption}
\usepackage{tabu}
\usepackage{balance}
\usepackage{footnote}
\makesavenoteenv{tabular}
\makesavenoteenv{table}
\usepackage{tikz}
\usepackage{adjustbox}
\usepackage[edges]{forest}
\usepackage{pifont}

\usepackage{fontawesome}

\def\eqref#1{equation~\ref{#1}}

\def\1{\bm{1}}

\DeclareMathAlphabet{\mathsfit}{\encodingdefault}{\sfdefault}{m}{sl}
\SetMathAlphabet{\mathsfit}{bold}{\encodingdefault}{\sfdefault}{bx}{n}

\DeclareMathOperator*{\bertdot}{BERT_\text{DOT}}
\DeclareMathOperator*{\bert}{BERT}

\usepackage{array, booktabs}
\usepackage{graphicx}

\usepackage{pifont}
\usepackage[noabbrev]{cleveref}
\usepackage{changepage}

\definecolor{niceGreen1}{RGB}{41, 135, 66}
\definecolor{niceGreen2}{RGB}{147, 184, 103}
\definecolor{niceYellow2}{RGB}{189, 164, 81}
\definecolor{niceRed2}{RGB}{186, 56, 65}
\definecolor{niceBlue}{RGB}{82, 82, 179}

\copyrightyear{2022} 
\acmYear{2022} 
\setcopyright{acmlicensed}

\newcommand{\cmark}{\textcolor{niceGreen2}{\ding{51}}}\newcommand{\xmark}{\textcolor{niceRed2}{\ding{55}}}\newcommand{\appr}{\textcolor{niceBlue}{$\pmb{\approx}$}}

\newcommand{\eg}{\emph{e.g.}}

\usepackage{xspace}
\newcommand{\cEffective}{{\texttt{C-Effective}}\xspace}
\newcommand{\cEfficient}{{\texttt{C-Efficient}}\xspace}
\newcommand{\cTradeoff}{{\texttt{C-Tradeoff}}\xspace}

\newcommand{\cRobust}{{\texttt{C-Robust}}\xspace}
\newcommand{\cRobustLength}{{\texttt{C-Length}}\xspace}
\newcommand{\cRobustFrequency}{{\texttt{C-Frequency}}\xspace}
\newcommand{\cRobustLexical}{{\texttt{C-Lexical}}\xspace}
\newcommand{\cRobustMemory}{{\texttt{C-Memory}}\xspace}
\newcommand{\cRobustMargin}{{\texttt{C-Margin}}\xspace}

\newcommand{\cBias}{\textbf{\texttt{C-Bias}}\xspace}
\newcommand{\cMaintain}{\textbf{\texttt{C-Maintainability}}\xspace}

\newcommand{\cEnvironment}{\textbf{\texttt{C-Environment}}\xspace}

\author{Sebastian Hofst\"{a}tter}
\affiliation{  \institution{TU Wien}
 }
\email{s.hofstatter@tuwien.ac.at}

\author{Nick Craswell}
\affiliation{  \institution{Microsoft}
}
\email{nickcr@microsoft.com}

\author{Bhaskar Mitra}
\affiliation{  \institution{Microsoft}
}
\email{bmitra@microsoft.com}

\author{Hamed Zamani}
\affiliation{  \institution{University of Massachusetts Amherst}
}
\email{zamani@cs.umass.edu}

\author{Allan Hanbury}
\affiliation{  \institution{TU Wien}
}
\email{allan.hanbury@tuwien.ac.at}

\begin{document}

\title{Are We There Yet? A Decision Framework for Replacing~Term-Based~Retrieval~with~Dense~Retrieval~Systems}
\begin{abstract}

Recently, several dense retrieval (DR) models have demonstrated competitive performance to term-based retrieval that are ubiquitous in search systems. In contrast to term-based matching, DR projects queries and documents into a dense vector space and retrieves results via (approximate) nearest neighbor search. 
Deploying a new system, such as DR, inevitably involves tradeoffs in aspects of its performance. Established retrieval systems running at scale are usually well understood in terms of effectiveness and costs, such as query latency, indexing throughput, or storage requirements.
In this work, we propose a framework with a set of criteria that go beyond simple effectiveness measures to thoroughly compare two retrieval systems with the explicit goal of assessing the readiness of one system to replace the other. 
This includes careful tradeoff considerations between effectiveness and various cost factors. 
Furthermore, we describe \textit{guardrail} criteria, since even a system that is better on average may have systematic failures on a minority of queries. The guardrails check for failures on certain query characteristics and novel failure types that are only possible in dense retrieval systems. 

We demonstrate our decision framework on a Web ranking scenario. In that scenario, state-of-the-art DR models have surprisingly strong results, not only on average performance but passing an extensive set of guardrail tests, showing robustness on different query characteristics, lexical matching, generalization, and number of regressions. DR with approximate nearest neighbor search has comparable low query latency to term-based systems. The main reason to reject current DR models in this scenario is the cost of vectorization, which is much higher than the cost of building a traditional index. It is impossible to predict whether DR will become ubiquitous in the future, but one way this is possible is through repeated applications of decision processes such as the one presented here.

\end{abstract}
\keywords{Information Retrieval, Machine Learning, Neural Ranking}

\maketitle

\section{Introduction}

Term-based indexes comprise of a list of terms and their locations in a text collection.
The idea of term-based indexing predates modern computers and their application can be traced back to the Bible concordances of the $13^\text{th}$ century~\citep{fenlon1913catholic} which were verbal indexes to the Bible.
Subsequently, the oldest printed indexes appeared in the mid-$15^\text{th}$ century~\citep{wellisch1986oldest}.
Today, term-based indexes in the form of inverted-indexes~\citep{zobel2006inverted} are employed in most computational search systems, including commercial web search engines.

\citet{boytsov2016off} have argued that another approach involving nearest neighbour lookup should be revisited in the context of search.
In light of emerging neural representation learning models for retrieval~\citep{mitra2018introduction}, such an approach combined with learned dense vector representations---sometimes referred to as dense retrieval (DR)---have renewed interests in the research community.
Initial attempts at dense retrieval---both in the pre-deep learning era~\citep{hofmann1999probabilistic} and in the early days of the neural information retrieval~\citep{ganguly2015word, ai2016analysis}---suffered heavily from off-topic false positive matches under the full retrieval setting which typically then necessitated that dense retrieval be combined with term-based retrieval to achieve reasonable result quality.
However, recently there have been several significant improvements~\citep{karpukhin2020dense,  xiong2020approximate, ding2020rocketqa, Hofstaetter2021_tasb_dense_retrieval} in training methodologies for dense retrieval models leading to many top ranking runs on public benchmarks like MS MARCO~\citep{craswell2021ms} and various open-domain question answering (QA) datasets~\citep{karpukhin2020dense}.
In spite of achieving competitive performance with traditional term-based indexing on different research benchmarks, the question remains whether these dense retrieval systems are ready to replace existing term-based retrieval in practical search systems.
The answer to that question requires more than simply comparing these systems based on mean effectiveness metrics on a query workload, and involves careful consideration of tradeoffs between several different costs and effectiveness, as well as the result quality on subsets of the query distribution.

A critical dimension that is often overlooked or under-studied in TREC-style~\citep{trec2019overview,trec2020overview} and leaderboard~\citep{craswell2021ms} evaluations, is the cost of deploying these models.
In production retrieval systems, cost typically can be a combination of several factors---\eg, indexing cost, query processing cost, and even environmental impact of training deep models~\citep{bender2021dangers}---and should be measured under the exact conditions in which the system will be deployed.

After testing two or more systems, we may have many observations in terms of cost factors and effectiveness measures.
A rational choice for deployment of a system assumes that all choices are alongside the Pareto optimum of cost vs. effectiveness.
If a candidate system is not alongside the known Pareto optimum, it would naturally fall out of consideration immediately, because there are better options in all dimensions.
However, as soon as we compare systems along the Pareto optimum, we face tradeoff decisions and constraints.

When evaluating effectiveness of ranking models, be it at TREC or on a leaderboard, it is also common practice to compare and rank systems based on performance metrics averaged over a sample set of queries.
However, the mean value of the metric may unwittingly hide critical systemic failures and mislead us on the model's readiness for deployment.
For example, previous work~\cite{Mitra2017,Xiong2017} has argued for the need to incorporate lexical matching features in neural models to deal with rare terms, especially under the full retrieval setting~\cite{kuzi2020leveraging, mitra2021improving}.
It is therefore reasonable to question if dense retrieval methods on their own may systemically fail to retrieve relevant documents when the query contains rare terms.
The dense retriever may show other systemic weaknesses, such as under-performing on longer queries or failing to retrieve longer documents, given theoretical constraints on learning fixed sized embeddings for long text~\cite{luan2020sparse} or demonstrated proclivity of neural retrieval models towards over-retrieving shorter documents~\cite{Hofstaetter2020_sigir} when inspecting only the beginning of documents for efficiency reasons.
Dense retrieval models may fail on queries that are out of distribution~\cite{cohen2018cross,craswell2021ms,thakur2021beir} with respect to the training data.
Performance on out-of-distribution queries can be critical for real world search systems where data distributions can change over time, or vary across locations and user demographics.
It may even be important to distinguish between different types of failures---\eg, retrieving non-relevant but related documents as opposed to retrieving results that are completely off-topic.
Such embarrassing failures can have disproportionate impact on the system's brand and long term user engagement~\cite{Ovidiu2016_satisfaction}. 
In this work, we design a decision framework to systematically compare dense retrieval to term-based indexes to determine their suitability of deployment in real production search systems as a replacement for the latter.
Unlike previous work, \eg,~\cite{karpukhin2020dense}, we are not concerned about whether dense retrieval outperforms term-based retrieval across different benchmarks.
Instead, we are interested to study if dense retrieval can be deployed in the context of a given benchmark or application, and the various detailed considerations that influence that specific decision.
This paper has one full case study of applying the decision framework.
To almost our surprise, in our chosen test case, carefully-designed dense retrieval models outperform traditional term-based retrieval in almost every dimension, including analysis designed to highlight problems such as handling of rare terms. However, this is only one scenario and even here, the ultimate shipping decision is constrained by available hardware budget increases and the individual importance of different cost factors.

Designing a decision framework to compare two retrieval systems with the explicit purpose of determining whether one system should replace another is inherently challenging.
We note that our proposed framework is general and appropriate for comparing arbitrary search systems.
So, in addition to highlighting the efficacy of state-of-the-art dense retrieval methods over traditional search using inverted-indexes, an important contribution of this work is the framework itself.
We posit that the proposed framework can guide rational and informed choices between different search architectures and models, albeit, individual cases require individual weighting decisions of which cost is more important.

Our main contributions and findings are summarized as follows:
\begin{itemize}[leftmargin=*]
    \item We propose a set of criteria to compare retrieval systems with cost-effectiveness tradeoff considerations and guardrails to ensure robustness against failures.
    \item We propose a general decision framework to guide a decision maker in applying our criteria.
    \item We implement our framework on one scenario, finding state-of-the-art DR models are ready to replace term-based systems, unless cost of vectorization is the overriding concern.
\end{itemize}

\section{Related Work}

\paragraph{\textbf{Neural Ranking and Dense Retrieval}}
Neural ranking models (NRMs) have been widely studied in the past few years and have led to substantial and significant improvements in terms of retrieval effectiveness~\cite{mitra2018introduction,Guo2020NRMSurvey}. Following learning to rank models and their applications in multi-stage cascaded retrieval systems, NRMs are mostly designed to re-rank a small number of documents retrieved by one or more efficient early-stage retrieval models, such as BM25~\cite{robertson1995okapi}. Notable models of this category include DRMM~\cite{Guo2016}, Duet~\cite{Mitra2017}, KNRM~\cite{Xiong2017}, TK~\cite{Hofstaetter2019_ecir}, and BERT re-ranking~\cite{nogueira2019passage}. However, the performance of these models is bounded by the recall of the early stage retrieval models. To address this shortcoming, \citet{zamani2018neural} introduced SNRM, the first standalone neural ranking model that can retrieve documents directly from a large-scale collection.
Recently, models use dense representations obtained from pre-trained contextual language models, e.g., BERT~\cite{devlin2018bert}. 
They are often called \textit{dense retrieval} models. ColBERT~\cite{khattab2020colbert}, ANCE~\cite{xiong2020approximate}, RocketQA~\cite{ding2020rocketqa} and TAS-Balanced~\cite{Hofstaetter2021_tasb_dense_retrieval} are among notable dense retrieval models. These dense retrieval models have been recently used in the literature as state-of-the-art NRMs, however, we believe that they have never been appropriately evaluated and studied. Prior work on dense retrieval~\cite{Prakash2021Dense,Wang2021Dense,xiong2020approximate,Hofstaetter2021_tasb_dense_retrieval} mostly report the average retrieval performance on a query set. Therefore, it is still an open question to what extent these dense retrieval models can replace the established and robust term matching models that use inverted indexes for efficient retrieval. This paper introduces a comprehensive evaluation framework, which can be applied to make the replacement decision in a given scenario. 

\paragraph{\textbf{Axiomatic Analysis}}
The proposed evaluation framework is slightly related to the axiomatic analysis literature. Axiomatic analysis is a study of retrieval models using a set of well-defined and easy-to-measure constraints (called axioms) and the intuition is that every model should satisfy those axioms. Therefore, axiomatic analysis can provide guidelines for further development of models by highlighting the unsatisfied axioms. \citet{Fang2004Axiomatic} introduced axiomatic analysis to information retrieval with the goal of improving term-based retrieval models, such as BM25. Their approach has been further extended to a wide range of models, such as pseudo-relevance feedback \cite{Montazeralghaem2016AxiomaticPRF,Clinchant2013AxiomaticPRF} and query performance prediction~\cite{Makarenkov2015AxiomaticQPP}. It has also been employed for improving neural ranking models~\cite{Rosset2019AxiomaticNRM,Rennings2019AxiomaticNRM}. Moreover, \citet{Busin2013Axiometrics} went beyond retrieval models and brought axiomatic analysis to study evaluation metrics, called axiometrics. They suggest a set of axioms that a metric should satisfy. Similar to this body of work, our evaluation framework also consists of a number of criteria or axioms. We believe that making decisions on the use of dense retrieval in search engines requires analysis of multiple proposed criteria. The criteria we introduce in this paper are novel and unlike prior work on axiomatic analysis, some of the criteria we propose may introduce a tradeoff and system designers should make decision based on their needs. 

\paragraph{\textbf{Efficiency-Effectiveness Tradeoff in IR}}
IR systems mainly aim at \textit{retrieving relevant documents} from large-scale collections, \textit{efficiently}. Even though published research mostly focuses on either effectiveness or efficiency considerations, taking both of them into account is at the heart of IR systems. There have been numerous efforts for developing and evaluating IR systems from both of these perspectives~\cite{Croft2009SEIRiP,zobel2006inverted,EfficientQAWorkshopReport}. For instance, \citet{Asadi2013} studied efficiency-effectiveness tradeoffs in candidate generation for multi-stage cascaded retrieval systems. Later on, \citet{Clarke2016} extended this work to end-to-end multi-stage systems and developed a model that does not require relevance judgement information. Refer to the tutorial by \citet{Lucchese2017} on the tradeoffs of multi-stage cascaded systems for more detail. 
Optimizing retrieval systems by considering both effectiveness and efficiency measures is a multi-objective optimization problem. A natural approach to address this optimization problem is to use the Pareto frontier~\cite{Hsiao2015ParetoImage,Mackenzie2020ParetoAdditivity,Ma2021}. 

Unlike previous work in this area which mainly focused on multi-stage cascaded systems, we study dense retrieval as a standalone retrieval model. The efficiency-effectiveness tradeoff in dense retrieval models is relatively unknown and this work will provide suggestions on the practical use of dense retrieval in search engines.

\section{Decision criteria}
\label{sec:criteria}

Our goal is to decide whether to replace an old system with a new system. Our case study in this paper considers dense retrieval as the new system with traditional indexing as the old system. This would mean replacing a well-studied and proven retrieval system with a promising, but less understood retrieval system. Dense retrieval has some known potential `deal breakers', which could make it unusable in practice. The cost of vectorization and nearest-neighbor search may be too high. Despite good average effectiveness, there may be subsets of queries where results are extremely bad, perhaps when dealing with rare query terms that were not seen during training. If there is data drift after deployment, the improvements in performance may disappear. To handle these concerns, as in any given application, we suggest identifying a set of application-specific decision criteria.

\subsection{Overall Criteria}

The starting point when comparing retrieval systems is the mean effectiveness \citep{karpukhin2020dense,  xiong2020approximate, ding2020rocketqa, Hofstaetter2021_tasb_dense_retrieval} on a dataset that matches the target application. We apply statistical tests to determine if there is a significant difference in means, adopting the following notation.

The operator $\underset{sig.T}{\gg}$ means statistically significantly larger, as tested with a test $T$, for two results $X$ \& $Y$:
\begin{equation}
\begin{aligned}
X \underset{sig.T}{\gg}  Y
\end{aligned}
\label{eq:c1-main}
\end{equation}

\begin{itemize}[leftmargin=1.9cm]
    \item[\textbf{\cEffective}] \textit{Using a metric (f.e. NDCG@10) the new system B should be significantly better than system A: $B \underset{sig.T}{\gg}  A$.} 
\end{itemize}

We can also require a margin of improvement, to eliminate tiny but significant improvements, which may be possible in some cases.

Among our overall decision criteria, some could require a significance test, as in Eq.~\ref{eq:c1-main}. However, other quantities are less suitable for significance testing, such as the size of the index. If the new system requires a much larger index, this can be captured by an absolute threshold (the index size must not exceed 20GB in production) or ratio (the index shall not grow by a factor of five). Such limits, defined by the decision maker, may be of great practical importance, depending on the scenario. It is also possible to define a criterion that combines statistical significance and practical significance, to discount small-but-significant improvements, that may not be worth deploying.

Our decision framework in Section~\ref{sec:decision} tells us how to make a decision based on multiple criteria. Still, the cost of deploying the system may include a variety of different aspects, such as query latency, indexing throughput, and storage requirements. The decision maker may decide that a particular cost factor is the most important one, for example that query latency is important, but indexing costs can be ignored. Another option is to apply a comparative transform $\Phi$ to each cost factor, to put them on the same scale, and use a weighted combination to summarize the cost factors.  

The comparative transform $\Phi$ can take many forms, involving any monotonic function, but a simple version is a scaled fraction comparing the new approach's performance $x$ to baseline performance $y$ with importance weight $\alpha$:
\begin{equation}
\begin{aligned}
\Phi(x,y) = \frac{x}{y} * \alpha
\end{aligned}
\end{equation}
By transforming several cost measurements and summing, the decision maker can summarize several cost factors as a single number.

\begin{itemize}[leftmargin=1.9cm]
    \item[\textbf{\cEfficient}] \textit{The cost of the new model B should not exceed the cost of the old model A by more than a certain factor ($N$ times) or more than a certain margin ($D$ distance).}
\end{itemize}

This can be applied on individual cost factors or a transformed and combined cost factor.
Once the decision maker has chosen application-specific criteria for overall \cEffective and \cEfficient, they can consider some criteria that do not measure overall performance: robustness criteria.

\subsection{Robustness Criteria}

Overall improvements in effectiveness, particularly when some query performances become a lot better, can hide systematic problems in smaller sub-groups of queries. For example, a system that is better overall might be much worse at handling queries with rare words. This could be a \textit{deal breaker}, if this makes it impossible to retrieve certain content. Users may notice that they can not search for a person by name, if the person's name has rare words that were not seen during model training, and they may reject the new system outright.

\begin{itemize}[leftmargin=1.5cm]
    \item[\textbf{\cRobust}] \textit{The new system B should not exhibit systematic failure patterns compared to system A; which might be hidden in the aggregated metrics.}
\end{itemize}

Observing the result metric $R$ for a certain subset of model independently selected queries $\hat{Q}$, there should be no statistically significant loss:
\begin{equation}
\begin{aligned}
\texttt{NOT} \left( {R}_{B}(\hat{Q}) \underset{sig.T}{\ll} {R}_{A}(\hat{Q}) \right)
\label{eq:c2-main}
\end{aligned}
\end{equation}

Now, we are able to define various subsets of our query distribution to study these systematic model differences. 
The new system B should not categorically fail on specific query characteristics, such as length  or term rarity, even if those queries are underrepresented in the data.

In IR we have a rich history of studying query characteristics, common dimensions include the query token length. For \textbf{\cRobustLength} we define a set of queries with specific token length between $m$ and $n$ as:
\begin{equation}
\begin{aligned}
\hat{Q} = \{q \ | \ m < len(q) < n , q \in Q\}
\label{eq:cRobustLen}
\end{aligned}
\end{equation}
Another common categorization of queries is to utilize the query token frequency. We define \textbf{\cRobustFrequency} as a set of queries with specific minimum token frequency $TF$ between $m$ and $n$ as:
\begin{equation}
\begin{aligned}
\hat{Q} = \{q \ | \ m < \min TF(q) < n , q \in Q\}
\label{eq:cRobustTF}
\end{aligned}
\end{equation}

Selecting regions of the query distribution is not limited to these two examples. Any slice of queries is possible, albeit it is important to select them system independently. If the selection depends directly or indirectly on the quality of one of the participating systems, the criterion in Eq. \ref{eq:c2-main} losses its expressiveness.

\begin{itemize}[leftmargin=1.5cm]
    \item[\textbf{\cRobustLexical}] \textit{Novel types of possible failures from a new system B need to be specifically tested to show that B is robust.}
\end{itemize}

In the case of DR models, one such novel failure type can be determined by the subset of queries, which contains a query if the top ranked (up to $r$) passages $p$ contain only up to $n$ token overlaps (as defined by the set intersection $\cap$): 
\begin{equation}
\begin{aligned} 
\hat{Q} = \big\{q \ \big\vert \ & n = |q \cap p| , rank(q,p) < r , p \in P , q \in Q \big\}
\label{eq:cRobustLexical}
\end{aligned}
\end{equation}

\begin{itemize}[leftmargin=1.5cm]
    \item[\textbf{\cRobustMemory}] \textit{If a system relies on machine learning it should be able to handle the open nature of the retrieval problem: Out of distribution query types and topics during inference.}
\end{itemize}

Given a measure of query similarity $C$, we can determine for every training and evaluation query their distance, to observe if very different queries in the evaluation set $Q$ (here $Q_{Eval}$ for readability) , as defined by the threshold $\epsilon$, to the training set $Q_{Train}$, still provide similar retrieval performance for the set:
\begin{equation}
\begin{aligned}
\hat{Q} = \{q \ \big| \  C(q_T, q) > \epsilon , q \in Q_{Eval} , q_T \in Q_{Train}\}
\end{aligned}
\end{equation}
The query similarity $C$ can range from simple word count overlaps to more sophisticated vector semantic models. A requirement is to incorporate the two query sequences (with potential statistics about all queries $Q$) and output a single value measuring the similarity.

\cRobust is not limited to our defined sets (Eq. \ref{eq:cRobustLen}, \ref{eq:cRobustTF}, \ref{eq:cRobustLexical}), as many other approaches fit into its definition. \citet{mackie2021dlhard} create a special set of hard queries from the TREC DL Track '19 \& '20, with a combination of automatic and manual hardness classification.

\begin{itemize}[leftmargin=1.5cm]
    \item[\textbf{\cRobustMargin}] \textit{The new system B should only regress results on queries up to a threshold compared to the old system A.}
\end{itemize}

Formally defined, as the threshold $t$ between the result metric with at least a distance margin $\delta$ per query of the old system ${R}_{A}(q)$ and new system ${R}_{B}(q)$ over all queries $Q$ as:

\begin{equation}
\begin{aligned}
    t \geq \frac{\big|\{{R}_{A}(q) - {R}_{B}(q) \geq \delta \ , \ q \in Q \}\big|}{|Q|} 
    \label{eq:robust-margin}
\end{aligned}
\end{equation}
We define a minimum result change margin to be able to focus on embarrassing failures and not small differences explainable by noise in the evaluation. This is closely related to Robustness Index (RI)~\cite{Collins2009RI}.

\paragraph{\textbf{Additional Important Criteria.}} Even though the mentioned criteria are extensive, they may not be fully sufficient for the full consideration of deploying a system. In Section~\ref{sec:additional_criteria}, we review a number of additional criteria that decision makers should take into account.

\section{Decision Framework}
\label{sec:decision}

After the decision maker has chosen a set of criteria that are suitable for their application, and gathered observations against the criteria, they can apply our decision framework. We would first ask the decision maker to classify each criterion as primary or secondary. Among the primary criteria, we need to see at least one significant improvement to justify launching a new system. A secondary criterion is a `guardrail' or `deal breaker', where we do not require an improvement but a significant regression would cause us to decide against the new system.

Although we cannot tell the decision maker which criteria to use, we believe our proposed dense retrieval criteria are comprehensive, probing for known weaknesses of dense retrieval systems. Other dense retrieval studies could apply the same criteria, but with their own query workload and corpus. For other kinds of information retrieval deployment decision, the decision maker can choose criteria that are appropriate for their setting.

Our decision framework has a \textit{significance rule} that takes into account primary and secondary criteria, and a \textit{Pareto rule} that focuses on tradeoffs between primary criteria. To be deployed, a new system must satisfy both rules.

\paragraph{\textbf{Significance Rule.}} We summarize the results for each criterion as a win/tie/loss for the new system, denoted by \cmark/\appr/\xmark. This is determined via statistical significance tests (e.g. improved mean NDCG) and/or tests of practical significance (e.g., $0.01$ NDCG improvement, or $20\%$ query latency reduction). We strongly suggest using statistical significance tests for measurement of system effectiveness, but not for a criterion that is a single number, such as index size. The practical significance test can check whether the index size has grown above some limit (\xmark). We can also create a combined criterion, which requires both a statistically significant gain and sufficient magnitude of gain (\cmark). In this case, a small but statistically significant gain would be considered \appr.

Given the per-criterion outcome \cmark/\appr/\xmark, the significance rule considers primary and secondary criteria. For primary criteria, there should be some improvement (\cmark), to make the change worthwhile, and no loss (\xmark). For example, a statistically significant improvement in NDCG could be enough to satisfy this rule. For secondary criteria, we are looking at `deal breaker' or `guard rail' cases. We do not require a win but there should be no losses (\xmark). In summary, a system passes the significance rule if it has an improvement on a primary criterion, and no losses on any primary or secondary criteria.

\paragraph{\textbf{Pareto Rule.}} The Pareto rule considers tradeoffs between criteria without considering significance, making it complementary to the significance rule. Under Pareto analysis, we can consider the old and new system, but can also consider several parameterizations of the new system. In Pareto analysis, if system B is better on one criterion than system A, without being worse on any other, then B is a Pareto improvement over A, so we do not need to consider A. If no system is better than system B on any criterion without being worse on some other, then B is Pareto optimal. The set of Pareto optimal systems is the Pareto frontier.

For example, if a system has both higher NDCG and lower query latency than another system, then we can discard the latter system, if we are only considering those two criteria. If one system has higher NDCG and lower latency, while the other has higher latency and lower NDCG, then we should not discard either. Both are on the Pareto frontier.

If the old system is not on the Pareto frontier, it is Pareto dominated by the new systems. In any case, we should choose a Pareto optimal system, since otherwise it is possible to do better on a criterion without doing worse on any others. One way of doing so is to transform each criterion into a comparable scale, by selecting an appropriate monotonic transform $\Phi$, then we can simply choose the alternative furthest from the origin. A related option is to consider the change in total cost of running the search system, and trade this off against the magnitude of the NDCG improvement. This requires the decision maker to have some notion of the value of an NDCG improvement, but having some feel for this is probably fundamental to making any decision related to search quality, so we leave it to the decision maker to consider the return on investment in their application setting.

\section{Intended Audience}

The decision criteria and framework proposed in this work can be helpful for various stakeholders in the field of information retrieval. Without the loss of generality, we describe some common uses:

\noindent\textbf{\textit{Practitioners}} could adapt our framework and decision criteria to their specific setting and use it to guide them, whether they should consider deploying a neural retrieval system in place or in addition of a traditional term-based retrieval system. 

\noindent\textbf{\textit{Paper reviewers}} could use our framework as a guide to critically inspect novel claims, especially if these claims only focus on either the cost benefits or the effectiveness of a new method. 

\noindent\textbf{\textit{Paper authors}} may use the full or parts of our framework to evaluate their proposed methods. Our robustness criteria can function as a strong secondary evaluation aspect for novel retrieval systems, if the main overall cost-effectiveness remains inconclusive.

\section{Case Study}

In this section, we showcase our framework in a case study on dense retrieval models using MS MARCO and TREC Deep Learning Tracks data. Since we expect dense retrieval to have better search effectiveness but with higher cost, our primary criteria are NDCG@10 (requiring a statistically significant difference with a two-tailed paired t-test; $p < 0.05$; \cEffective \cmark) and an aggregated cost that combines query latency and indexing costs (Eq.~(\ref{eq:cost-aggregation})).

Making decisions about cost factors is not as general as effectiveness results. Different practitioners are likely constrained in different ranges, therefore we use our decision framework to showcase different decision makers, without making a general claim. Providing a single answer to the question \textit{Are we there yet (as a community)?}, however convenient is simply not possible, due to the diverse nature of the retrieval settings. However, we provide the tools to answer specific use cases and settings for the question: \textit{Are we there yet (in our setting)?} We carry out a Pareto analysis on our primary criteria, considering several alternative new systems for our specific use case.

We also consider several robustness criteria, which we count as secondary, where we are looking for failures rather than improvements (\cRobust \xmark). 

\paragraph{\textbf{The Dense Retrieval Model.}} 
In our experiments, we use the BERT$_\text{DOT}$ model as the dense retrieval system. It uses two independent $\bert$ computations (each time pooling the \texttt{CLS} vector output) to obtain the query ${q}_{1:m}$ and passage ${p}_{1:n}$ representations. It then computes the retrieval score based on the dot product similarity of the two representations:
\begin{equation}
\begin{aligned} 
& \vec{q} = \bert([\text{CLS};{q}_{1:m}])  \\
& \vec{p} = \bert([\text{CLS};{p}_{1:n}]) \\
& \bertdot({q}_{1:m},{p}_{1:n})  = \vec{q} \cdot \vec{p}
\end{aligned}
\end{equation}

This architecture decouples the costly encoding from the search. We can store every passage in an (approximate) nearest neighbor index $I$ for direct vector-based retrieval. The retrieval of the top $k$ hits for a given query $q$ is then formalized as:
\begin{equation}
\begin{aligned}
\operatorname{top}_{k} \big\{ \vec{q} \cdot \vec{p} \ \big| \ \vec{p} \in I \big\}
\end{aligned}
\end{equation}

In this study, we use the \textit{Standalone} and \textit{TAS-Balanced} trained instances of $\bertdot$, developed by \citet{Hofstaetter2021_tasb_dense_retrieval}. The \textit{Standalone} version is trained with binary relevance labels from MS MARCO \cite{msmarco16}. The \textit{TAS-Balanced} retriever is trained with pairwise and in-batch negative knowledge distillation using topic-aware sampling to compose batches. It is currently a state-of-the-art training technique for dense retrieval \cite{thakur2021beir}.

\paragraph{\textbf{Term-based Retrieval Baselines.}} We use BM25 \cite{robertson1995okapi} and the neural augmented indexes DeepCT \cite{dai2019contextaware}, DocT5query  \cite{nogueira2019doct5query}, and DeepImpact \cite{mallia2021learning} using both Lucene \cite{lin2021pyserini} as well as highly tuned results with PISA \cite{pisasearch} reported by \citet{mallia2021learning}. 

\paragraph{\textbf{Datasets.}} We conduct our experiments on the MS MARCO-v1 and TREC 2019-20 Deep Learning Track collections with 9 million passages and 3 million documents.  Following \citet{xiong2020approximate}, for the document collection we utilize two approaches: (1) taking only the first passage of a document (FirstP), and (2) using the maximum passage score as document score (MaxP).

\begin{figure}[t]
       \includegraphics[trim={0cm 0cm 0cm 0cm},width=0.6\textwidth]{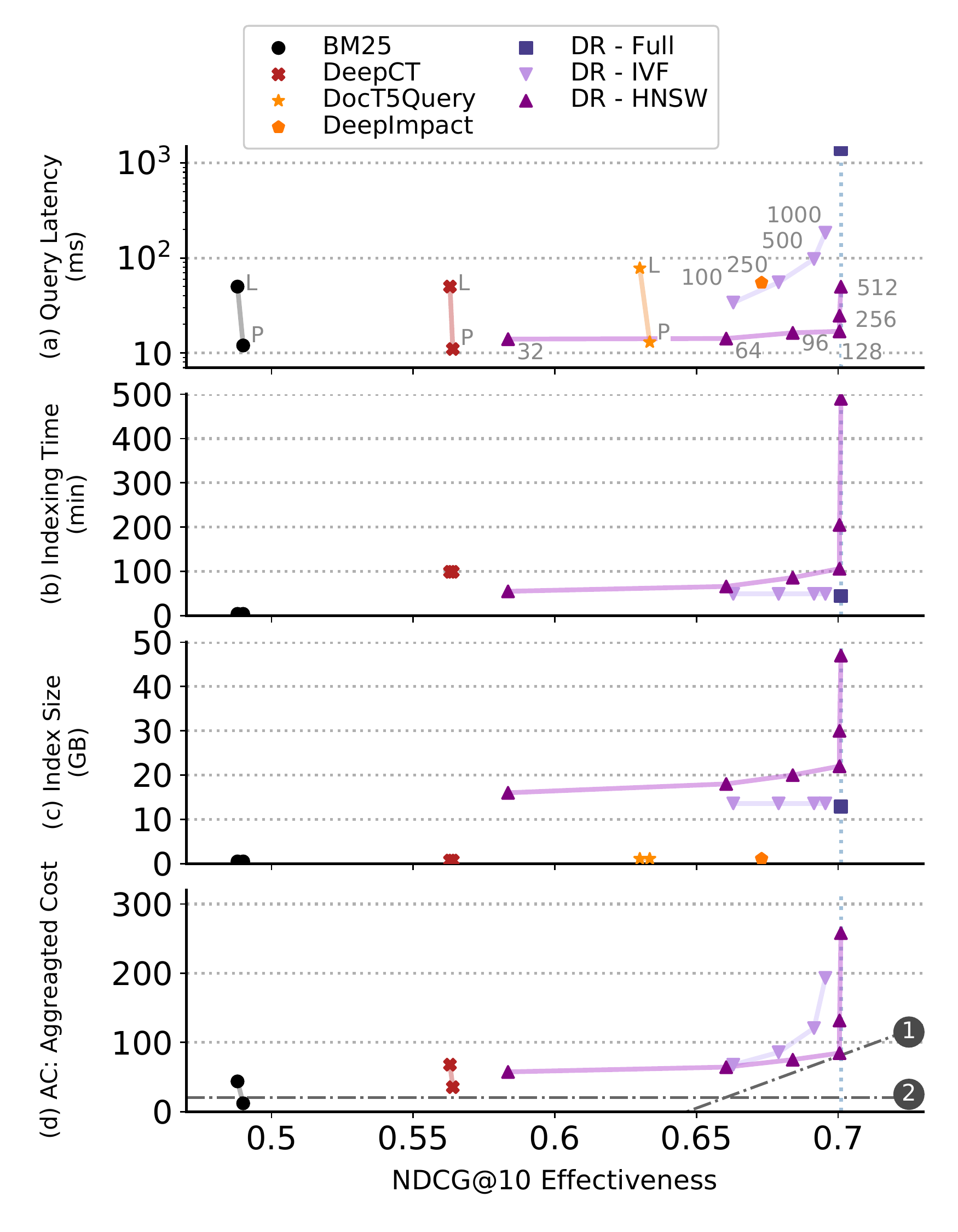}
    \centering
        \caption{TREC-DL passage retrieval comparison of the cost-effectiveness tradeoff for term-based and dense retrieval.}
    \label{fig:Pareto}
    \end{figure}

\subsection{Primary Criteria}
\label{sec:primary-criteria}

We first apply the decision framework to our primary criteria, NDCG@10 and aggregated cost.

\paragraph{\textbf{Study Goal.}} Identify a solution that has a significant NDCG@10 gain (significance rule) and the right tradeoff of NDCG@10 improvement and aggregated cost (Pareto rule). Since the tradeoff depends on the decision maker's priorities, we consider a decision maker that is willing to make a tradeoff (\ding{202}) and one that is very cost-sensitive (\ding{203}).

\paragraph{\textbf{Study Design.}} We utilize the mentioned TAS-Balanced trained dense retriever with three search approaches: (1) exhaustive search; and (2) the approximate nearest neighbor (ANN) method based on Hierarchical Navigable Small World graph (HNSW) \cite{malkov2018hnsw}; and (3) the ANN method based on Inverted Files (IVF) \cite{ivf_index2003}. We use the Faiss library \cite{faiss2017} for conducting our experiments.
We modulated HNSW and IVF with different hyper-parameter settings that guide the internal tradeoff between cost and quality (i.e., the number of graph-node neighbors and the number of lookup clusters). For measuring cost, we use encoding \& indexing time (using a GPU), storage requirement, and single average query latency (using a CPU). For measuring effectiveness, we use NDCG@10 averaged over the TREC-DL query sets. We conducted our experiments on a system with a 16 core Intel Xeon E5 and TITAN RTX GPU.

The aggregation of cost factors -- such as latency $l$, indexing time $i$, and storage $s$ -- is highly dependent on the scenario: whether we have a high query workload, a high document update frequency, or other requirements. Here, we instantiate our cost side of \cTradeoff with an exemplary balanced aggregated cost (AC) anchored to BM25 per method $m$: 
\begin{equation}
\begin{aligned}
AC_{m} = \frac{{l}_{m}}{{l}_{BM25}} * \alpha + \frac{{i}_{m}}{{i}_{BM25}} * \beta + \frac{{s}_{m}}{{s}_{BM25}} * \gamma
\end{aligned}
\label{eq:cost-aggregation}
\end{equation}
where $\alpha$, $\beta$, and $\gamma$ control the importance of each cost component. In Figure \ref{fig:Pareto}, we set $\alpha$ to 10, and $\beta$ \& $\gamma$ to 1. As this weighting has major impact on the decision and conclusion, we also provide alternative weighting results in Appendix \ref{sec:appendix-weighting}. 
We model a decision maker who cares much more about relative increases in response time, since these affect users directly, than they do about increases in indexing costs, perhaps because they already have sufficient available resources to handle a new system with moderate requirements.

\paragraph{\textbf{Study Analysis.}}

For the passage dataset, we show our  findings combined in Figure \ref{fig:Pareto}. Each subfigure shares the effectiveness on the x-axis. 
All improvements over BM25 are statistically significant (\cEffective \cmark), making many systems viable on that criterion. We will discuss cost factors then move on to our efficiency criterion.

On top in Figure \ref{fig:Pareto} (a) we compare the effectiveness with the log-scaled query latency for a single query (on CPUs). Query latency is one of the most common used cost metrics in IR and highly influential for user satisfaction \cite{kohavi2013online,zamani2018neural}. Here, we observe DR with HNSW approximation is Pareto dominant over other DR approaches. If we imagine a decision maker who mainly cares about the NDCG@10 and latency, they would be likely to select HNSW 128. 
If we instead look at indexing time and index storage requirements in Figure \ref{fig:Pareto} (b) and (c) respectively, HNSW exhibits much higher cost than a simple inverted index or the alternative ANN method IVF. Concurrently with this work, compression techniques have been proposed to reduce the required storage \cite{ma2021simple,zhan2021jointly}, however they also follow a tradeoff pattern, reducing their effectiveness as well. DocT5Query \& DeepImpact (which uses DocT5Query) are magnitudes slower (with 19,200 minutes) than all other approaches at indexing time, therefore they are not visible in Figure \ref{fig:Pareto} (b) \& (d). 
We show combined costs with our exemplary aggregation formula in Eq. \ref{eq:cost-aggregation} in Figure \ref{fig:Pareto} (d). Here, we added two decision lines (marked with \ding{202} and \ding{203}) for our two exemplary types of decision makers. Each line indicates a set of solutions that the decision maker would consider equally good, intersecting with the Pareto frontier. A decision maker that is willing to make some tradeoff \ding{202} would choose the bottom right point, which is HNSW-128 (\cEfficient  \cmark). However, if absolutely no cost increase is allowed, as in scenario \ding{203}, then all neural approaches would be rejected (\cEfficient \xmark).

For the document dataset, we are particularly interested in the recall of dense retrieval methods, as both the FirstP and MaxP approaches used cut off the text (at 512 and 4000 tokens respectively) and do not index every single word, as the term-based indexes do. Our results on the TREC-DL query sets in Table \ref{tab:doc_results} show a similar trend to our passage retrieval in Figure \ref{fig:Pareto}: Dense retrieval, both in full and approximate (HNSW with 128 neighbors) settings, outperforms term-based indexes in most cases. Looking at the recall, we observe that a MaxP approach is needed to substantially outperform term-based methods. However, MaxP naturally increases the storage requirements compared to FirstP, as more vectors need to be stored.

To conclude, in our application setting with MS MARCO data, dense retrieval models with the right ANN hyper-parameter choice can deliver both an equally low query latency and higher effectiveness than term-based methods. However, this comes at the cost of increased indexing and storage resource requirements. The decision to replace term-based retrieval with dense retrieval is therefore constrained based on individual budgets and constraints.
Next we consider secondary significance criteria.

\begin{table}[t]
    \centering
    \caption{Document retrieval results for TREC-DL query sets. \textit{Line \# superscript indicates stat.sig. improvement; paired t-test ($p < 0.05$). nDCG cutoff at 10, Recall at 100.}}
    \label{tab:doc_results}
        \setlength\tabcolsep{4pt}
        \begin{tabular}{cl!{\color{lightgray}\vrule}ll!{\color{lightgray}\vrule}ll!{\color{lightgray}\vrule}r}
       
       \toprule
       &\multirow{2}{*}{\textbf{Model}} & \multicolumn{2}{c!{\color{lightgray}\vrule}}{\textbf{TREC'19}} & \multicolumn{2}{c!{\color{lightgray}\vrule}}{\textbf{TREC'20}} & \textbf{Index} \\

       &&MRR&Rec.&nDCG&Rec. & \textbf{Size} \\
       \midrule
       \textcolor{gray}{1} & BM25 & .523 & .581 & .507 & .706 &  2.3 GB \\
       \textcolor{gray}{2} & DocT5Query & .597 & .599 & .589 & .759 & 2.5 GB\\
        \arrayrulecolor{lightgray}
       \midrule

       \textcolor{gray}{3} & DR+Full: FirstP & .630 & .556 & .598 & .705 & 5 GB \\
       \textcolor{gray}{4} & DR+Full: MaxP & \textbf{.636}\textcolor{gray}{$^{1}$} & \textbf{.610}\textcolor{gray}{$^{5}$} & \textbf{.639}\textcolor{gray}{$^{15}$} & .757\textcolor{gray}{$^{5}$}  & 10 GB \\
       \midrule
       \textcolor{gray}{5} & DR+HNSW: FirstP  & .607 & .542 & .586 & .684 & 8 GB \\
       \textcolor{gray}{6} & DR+HNSW: MaxP & .606 & .561 & .630\textcolor{gray}{$^{1}$} & \textbf{.760}\textcolor{gray}{$^{5}$} &  18 GB \\

        \arrayrulecolor{black}

       \bottomrule
        \end{tabular}
\end{table}

\subsection{Robustness by Query Characteristic}
\label{sec:robust-characteristics}

The mean effectiveness gains of dense retrieval (even with ANN search) compared to term-based approaches are large. DR could completely fail on a large subset of queries and still perform better than a baseline on average. However, this can be a practical dealbreaker. Therefore, we make use of our secondary guardrail criteria to evaluate retrieval models by query characteristics.

\paragraph{\textbf{Study Goal.}} We address the following question: \textit{Do DR models struggle on certain queries categorized by length or term frequency?}

\paragraph{\textbf{Study Design.}} We utilize the criterion \cRobust to compare approaches by query characteristic, specifically we make use of length based \cRobustLength and frequency based \cRobustFrequency. We evaluate term-based and exhaustive TAS-Balanced DR on the large MSMARCO-DEV-49K passage query set (with almost 49 thousand queries; distinct from the training set) in terms of MRR@10. The reason for switching from TREC-DL to MSMARCO-DEV for this experiment is that we benefit from using as many test queries as possible  to reduce noise with uncommon characteristics. Note that the TREC-DL data contains less than 100 queries.

\begin{figure}[t]
  \centering
  \begin{subfigure}[t]{0.49\textwidth}
    \centering
   \includegraphics[trim={0.2cm 0cm 0.1cm 0cm},width=0.99\textwidth]{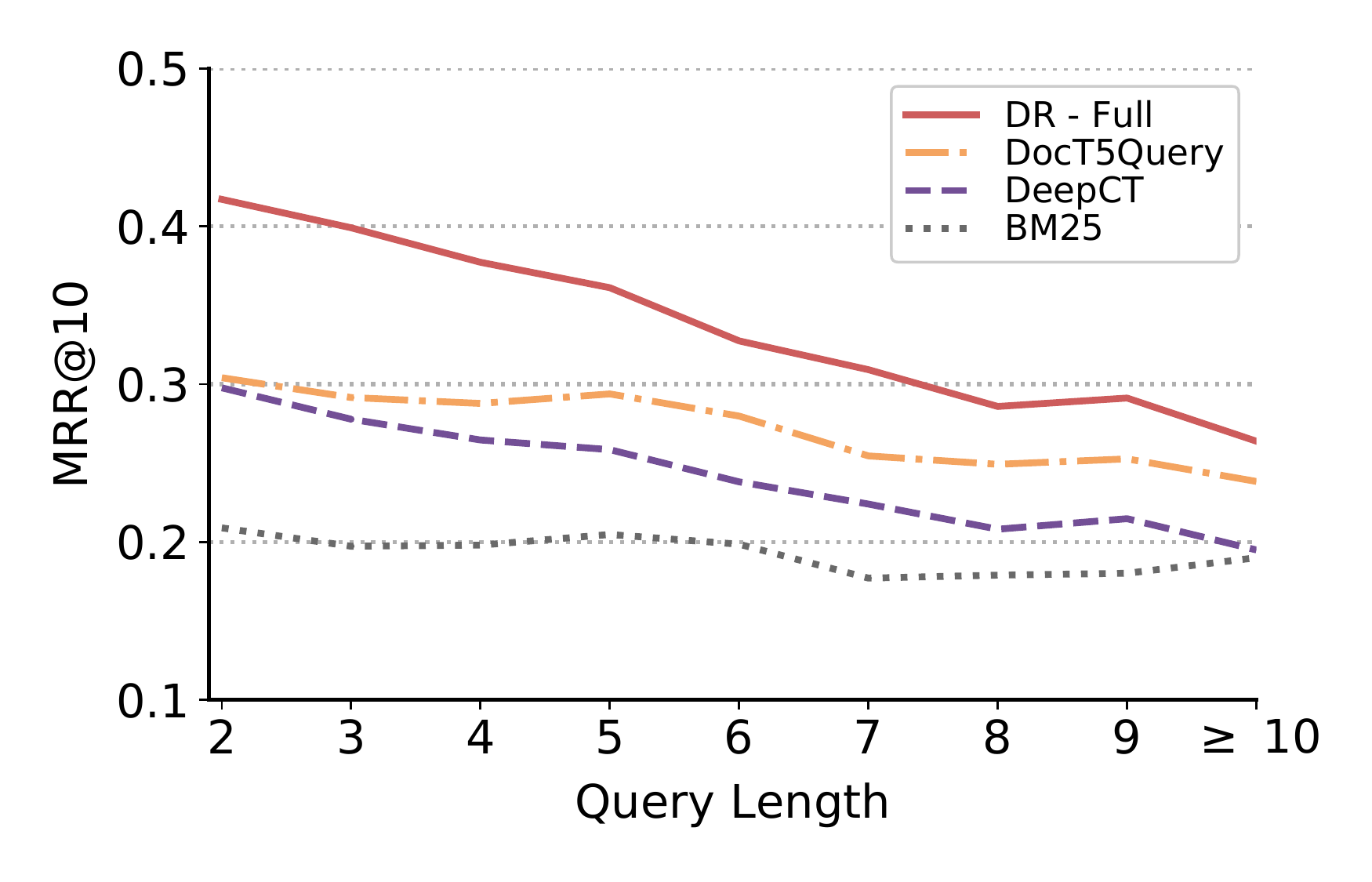}
    \centering
        \caption{By query length}
    \label{fig:qlen}
      \end{subfigure}        \begin{subfigure}[t]{0.49\textwidth}
    \centering
   \includegraphics[trim={0.2cm 0cm 0.1cm 0cm},width=0.99\textwidth]{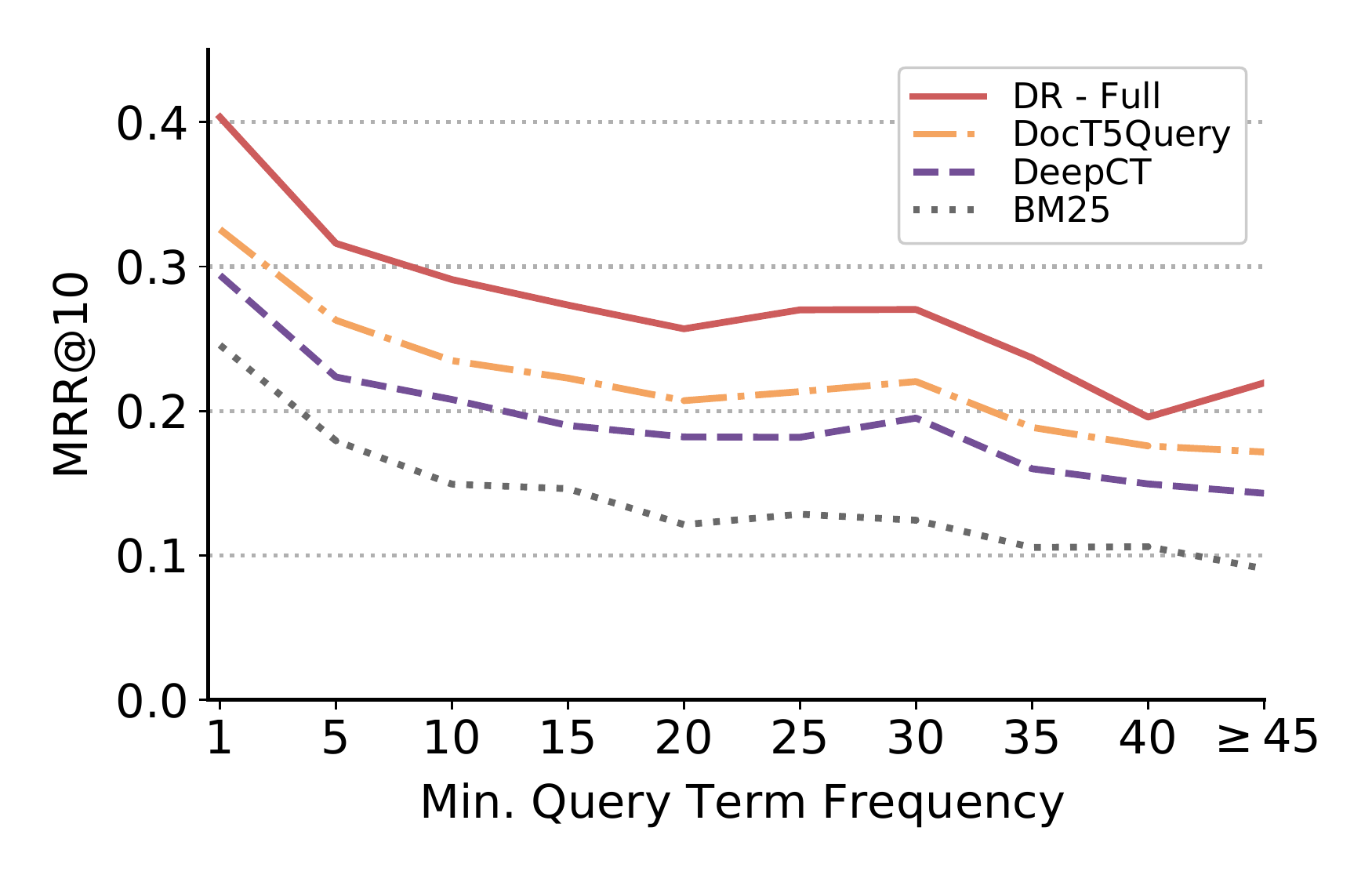}
    \centering
        \caption{By min. term frequency}
    \label{fig:min-qf}
      \end{subfigure}
      \caption{Comparison of MRR@10 effectiveness results on MSMARCO-DEV-49K by query characteristic.}
  \end{figure}

\paragraph{\textbf{Study Analysis.}}
We present the analysis by query length in Figure \ref{fig:qlen} and by minimum term frequency in Figure \ref{fig:min-qf}. In both cases, we observe dense retrieval performing better than term-based alternatives. This is especially surprising for queries with very rare terms (Figure \ref{fig:min-qf}). Even in cases where a query term is essentially unseen during training, the TAS-Balanced approach which does explicitly match terms is outperforming term-based approaches.
These are positive results showing that DR is consistent in the improvements and therefore DR does not fail our \cRobust tests (\appr).

\begin{table*}[t]
    \centering
    \caption{Annotation analysis of queries from MSMARCO-DEV-Large (49K queries) with zero or one subword overlap on the top-1 retrieved passage by the BERT$_\textbf{DOT}$ dense retrieval model. $\Delta$ P@1 shows the difference to the  DEV-7K set.}
    \label{tab:subword_analysis}
        \setlength\tabcolsep{3pt}
        \begin{tabular}{lr!{\color{lightgray}\vrule}rr!{\color{lightgray}\vrule}rr!{\color{lightgray}\vrule}rrl}
       
       \toprule
       \multirow{2}{*}{\textbf{Training}} & \textbf{Common} &  \multicolumn{2}{c!{\color{lightgray}\vrule}}{\textbf{Queries}} & \multicolumn{2}{c!{\color{lightgray}\vrule}}{\textbf{Query Tokens}} & \multicolumn{3}{c}{\textbf{Our Annotations}}  \\
       &\textbf{Tokens@1} & \# & \% & Total & Subwords & P@1 & $\Delta$ P@1 & 4-Graded Relevance Distribution \hspace{0.5cm} \\
       \midrule
       \multirow{2}{*}{None} & \textbf{0}  & 22,600 & 46.50 \% & 6.0 & 15\%  & .000 & --  & \hspace{-1.97cm}  \adjustbox{valign=t}{\begin{tikzpicture}[baseline] \tikz \fill [niceRed2] (0.,0.3) rectangle (5,0.); \node[text=white] at (-1.38,.15) {100\%}; \end{tikzpicture}} \\
                             & \textbf{1}  & 14,640 & 30.12 \% & 6.9 & 13\% & .000 & -- & \hspace{-1.97cm}  \adjustbox{valign=t}{\begin{tikzpicture}[baseline] \tikz \fill [niceRed2] (0.,0.3) rectangle (5,0.); \node[text=white] at (-1.38,.15) {100\%}; \end{tikzpicture}} \\

         \arrayrulecolor{lightgray}
         \midrule

       \multirow{2}{*}{Standalone} & \textbf{0}  & 163 & 0.33 \% & 5.0 & 16\%  & .313 & -.435  & \hspace{-1.9cm}  \adjustbox{valign=t}{\begin{tikzpicture}[baseline] \tikz \fill [niceGreen1] (0,0.3) rectangle (1.07,0.); \node[text=white] at (-0.535,.15) {21\%}; \tikz \fill [niceGreen2] (0.,0.3) rectangle (0.49,0.); \node[text=white] at (-0.245,.15) {10\%}; \tikz \fill [niceYellow2] (0.,0.3) rectangle (0.67,0.); \node[text=white] at (-0.335,.15) {13\%}; \tikz \fill [niceRed2] (0.,0.3) rectangle (2.76,0.); \node[text=white] at (-1.38,.15) {55\%}; \end{tikzpicture}} \\
                                   & \textbf{1}  & 1824 & 3.75 \% & 4.3 & 7\% & .580 & -.168 & \hspace{-1.17cm} \adjustbox{valign=t}{\begin{tikzpicture}[baseline] \tikz \fill [niceGreen1] (0,0.3) rectangle (2.25,0.); \node[text=white] at (-1.125,.15) {45\%}; \tikz \fill [niceGreen2] (0.,0.3) rectangle (0.66,0.); \node[text=white] at (-0.33,.15) {13\%}; \tikz \fill [niceYellow2] (0.,0.3) rectangle (0.8,0.); \node[text=white] at (-0.4,.15) {16\%}; \tikz \fill [niceRed2] (0.,0.3) rectangle (1.3,0.); \node[text=white] at (-0.65,.15) {26\%}; \end{tikzpicture}} \\

         \arrayrulecolor{lightgray}
         \midrule
       \multirow{2}{*}{TAS-Balanced} & \textbf{0}  & 37 & 0.01 \% & 4.0 & 8\% & .622 & -.173 &  \hspace{-1.2cm} \adjustbox{valign=t}{\begin{tikzpicture}[baseline] \tikz \fill [niceGreen1] (0,0.3) rectangle (2.16,0.); \node[text=white] at (-1.08,.15) {43\%}; \tikz \fill [niceGreen2] (0.,0.3) rectangle (0.95,0.); \node[text=white] at (-0.475,.15) {19\%}; \tikz \fill [niceYellow2] (0.,0.3) rectangle (0.54,0.); \node[text=white] at (-0.27,.15) {11\%}; \tikz \fill [niceRed2] (0.,0.3) rectangle (1.35,0.); \node[text=white] at (-0.675,.15) {27\%}; \end{tikzpicture}}  \\
                                   & \textbf{1}  & 1284 & 2.64 \% & 3.9 & 3\% & .715 & -.080 & \hspace{-.87cm} \adjustbox{valign=t}{\begin{tikzpicture}[baseline] \tikz \fill [niceGreen1] (0,0.3) rectangle (2.86,0.); \node[text=white] at (-1.43,.15) {57\%}; \tikz \fill [niceGreen2] (0.,0.3) rectangle (0.71,0.); \node[text=white] at (-0.355,.15) {14\%}; \tikz \fill [niceYellow2] (0.,0.3) rectangle (0.72,0.); \node[text=white] at (-0.36,.15) {14\%}; \tikz \fill [niceRed2] (0.,0.3) rectangle (0.71,0.); \node[text=white] at (-0.355,.15) {14\%}; \end{tikzpicture}}  \\

       \arrayrulecolor{black}
       \bottomrule
        \end{tabular}
\end{table*}

\subsection{Lexical Match Robustness}
\label{sec:lexical-matches}

Retrieving passages from an unconstrained vector space opens up a new failure scenario that term-based retrieval models cannot suffer from: returning passages without any term overlaps. 

\paragraph{\textbf{Study Goal.}} We aim at answering the following question: \textit{How well do different DR training approaches solve the lexical matching task?} We measure the extent of queries, where a DR model returns a top passage with low or no lexical text overlap. We constrain the problem to the first returned passage and the P@1 metric. For this case study, we instantiate our guardrail criteria \cRobustLexical.

\paragraph{\textbf{Study Design.}} 
We use Eq. \ref{eq:cRobustLexical} to select a set $\hat{Q}$ of queries, using the highest ranked passage by our dense retrieval models. We assume that cases without any ($n=0$) or only one ($n=1$) subword-token overlap between the query and the highest ranked passage represent hard-failures of missing lexical matches. 
Using exhaustive search, we compare three different methods: (1) a pre-trained DistilBERT model without fine-tuning; (2) a standalone fine-tuned model; and (3) the mentioned dense retrieval model based on TAS-Balanced. 
To receive robust results, we again conduct this study on the large MSMACRO-DEV-49K set. Furthermore, we manually judged all selected queries and their first retrieved passages (for the non-finetuned baseline we sampled 100 queries). The judgements were conducted by non-expert annotators on a 4-graded relevance scale, following the TREC-DL definition (Relevant: Perfect \& Partial; Non-Relevant: Topic \& Wrong). The annotators had to select relevant text spans for the two relevant classes, reducing the chance of inadvertent false positives.
With this we can confidently draw conclusions from this study that would not be possible with the noisy incomplete relevance judgements from MS MARCO.

\paragraph{\textbf{Study Analysis.}} In Table \ref{tab:subword_analysis}, we report the resulting query set sizes as well as our annotation results. First, we observe that our pre-trained only baseline completely fails - returning an irrelevant top-1 result with zero or one matching token for the majority of queries. This shows that there is indeed no guarantee that dense retrieval can do token matching or term (token n-gram) matching. However, when we turn towards the fine-tuned models with relevance data, we see that the standalone training already reduces the ratio of queries without matches to $.33\%$. The state-of-the-art TAS-Balanced retriever reduces this number further to $.01\%$ and the number of queries with a single match to $2.64\%$. Our annotations show that the true failures of these query sets are even smaller, because the P@1, while lower than the average P@1 of a random query sample, is still above $.622$ (no common tokens) and $.715$ (1 common token) for TAS-Balanced. 
To conclude, we only observe a practically insignificant fraction of queries where a missing term match is the cause of a non-relevant top retrieved passage, therefore TAS-Balanced does pass our \cRobust tests (\appr).

\begin{table}[t]
    \centering
    \caption{Training and test set MRR@10 results for TAS-Balanced using different training cluster-splits. \textit{Line \# superscript indicates stat.sig. improvement; paired t-test ($p < 0.05$).}}
    \label{tab:memorization-clusters}
        \setlength\tabcolsep{6pt}
        \begin{tabular}{clr!{\color{lightgray}\vrule}ll!{\color{lightgray}\vrule}ll}
       
       \toprule
       &\textbf{Training} & \multirow{2}{*}{\textbf{Ratio}} &  \multicolumn{2}{c!{\color{lightgray}\vrule}}{\textbf{Train (400K)}} & \multicolumn{2}{c}{\textbf{Test (49K)}}  \\
       &\textbf{Data} && All & C-Subset & All & C-Subset \\
       \midrule
       \textcolor{gray}{1} & BM25 & -- & .172 & .173 & .194 & .190 \\    
       \arrayrulecolor{lightgray}
       \midrule
       \textcolor{gray}{2} & All & 1.0    & \textbf{.345}\textcolor{gray}{$^{1345}$} & .346\textcolor{gray}{$^{135}$} & \textbf{.340}\textcolor{gray}{$^{134}$} & \textbf{.338}\textcolor{gray}{$^{134}$} \\
       \textcolor{gray}{3} & Uniform Reduction & 0.1    & .299\textcolor{gray}{$^{14}$} & .300\textcolor{gray}{$^{1}$} & .315\textcolor{gray}{$^{14}$} & .310\textcolor{gray}{$^{1}$} \\
       \arrayrulecolor{lightgray}
       \midrule
       \textcolor{gray}{4} & Subset-Clusters & 0.1    & .280\textcolor{gray}{$^{1}$} & \textbf{.429}\textcolor{gray}{$^{1235}$} & .298\textcolor{gray}{$^{1}$} & .321\textcolor{gray}{$^{13}$} \\
       \textcolor{gray}{5} & Non-Subset-Clusters & 0.9    & .343\textcolor{gray}{$^{134}$} & .302\textcolor{gray}{$^{1}$} & .339\textcolor{gray}{$^{134}$} & .334\textcolor{gray}{$^{134}$} \\

       \arrayrulecolor{black}

       \bottomrule
        \end{tabular}
\end{table}

\subsection{Memorization vs. Generalization}
\label{sec:memorization}

In contrast to many other traditional retrieval models which only have a few hyper-parameters to tune on a collection, dense retrieval models contain millions of parameters and require large-scale training data. This opens the possibility that DR models just memorize the query distribution (even if the test queries are technically distinct from the training set) and what looks like generalization across different query sets is actually memorization, as posited by us in \cRobustMemory.

\paragraph{\textbf{Study Goal.}} We aim to observe the robustness of dense retrieval to reduced training data and answer the question: \textit{Does the effectiveness of DR models come from memorization or a mix with generalization capabilities?}

\paragraph{\textbf{Study Design.}} We utilize the topic-cluster-guided training of TAS-Balanced as query-similarity measure to reduce the training to different cluster subsets. Queries are assigned a cluster using a baseline DR representation with dot-product similarity. We compare the following training splits in Table \ref{tab:memorization-clusters}: training on all 2000 query clusters (row 2); training on 10\% randomly selected queries, but using all clusters (row 3); and then our cluster-subsets with training on 10\% of clusters (row 4), as well as training on the remaining 90\% (row 5). We then evaluate these model instances on the training and the test set (both split in all available queries and queries associated with the 10\% subset clusters). 

\paragraph{\textbf{Study Analysis.}}

In Table \ref{tab:memorization-clusters} we see in rows 1, 2, \& 3 that BM25 and training on all clusters produces stable results with low margins between all and c-subset for both training and testing. As before we observe DR to vastly outperform BM25, albeit slightly reduced when we only train on 10\% of the training data. Even the strongly reduced-training DR instances pass our \cRobust tests (\cmark) compared to BM25.
If we only train on a 10\% cluster-subset (row 4) we see that it strongly memorizes the training set, outperforming the other DR methods substantially on the trained clusters. But on the test set it is outperformed on the same query clusters by the full training (row 2). Similarly, the 10\% cluster-subset training is also outperformed by more training data that does not contain queries from the evaluated clusters (row 5). 
To conclude, we do observe a tendency to memorize the training set, if the number of queries is small. A 10\% cluster-subset would fail our \cRobust tests (\xmark) compared to the other DR instances, but not to BM25 (\cmark). A 10\% drift in the query distribution from train to test (row 5) is as robust as no drift at all (\appr).

\subsection{Per Query Robustness of Improvement}
\label{sec:per-query-robustness}

Commonly in IR, the wins and losses on a query level between two competitive systems will always have some queries on both sides. The important factor is how many losses are tolerated, as we set in \cRobustMargin. Once a system is over a given threshold of missed queries, we need to reconsider deployment.

\paragraph{\textbf{Study Goal.}} In the previous studies (Sec. \ref{sec:robust-characteristics}, \ref{sec:lexical-matches}, \ref{sec:memorization}) we sliced dense retrieval results by query characteristics and effectiveness independent result properties. Now, we start our study from the other side and select queries using evaluation metrics. We want to answer: \textit{How large is the query set with the highest failure margin between DR and term-based models?}

\begin{table}[t]
    \centering
    \caption{Failure analysis of queries selected from MSMARCO-DEV (49K queries) with BM25 having an RR=1 and dense models RR=0 on sparse judgements. We annotated the top-1 retrieved result from the dense models.}
    \label{tab:bert_vs_bm25}
        \setlength\tabcolsep{6pt}
        \begin{tabular}{l!{\color{lightgray}\vrule}rr!{\color{lightgray}\vrule}rl}
       
       \toprule
       \multirow{2}{*}{\textbf{Training}} &  \multicolumn{2}{c!{\color{lightgray}\vrule}}{\textbf{Queries}} &  \multicolumn{2}{c}{\textbf{Our Annotations}}  \\
       & \# & \% &  P@1 & 4-Graded Relevance Distribution \hspace{0.5cm} \\
       \midrule
       Standalone &  911 & 1.87\% & .528 & \hspace{-.92cm} \adjustbox{valign=t}{\begin{tikzpicture}[baseline] \tikz \fill [niceGreen1] (0,0.3) rectangle (1.86,0.); \node[text=white] at (-0.93,.15) {37\%}; \tikz \fill [niceGreen2] (0.,0.3) rectangle (0.78,0.); \node[text=white] at (-0.39,.15) {16\%}; \tikz \fill [niceYellow2] (0.,0.3) rectangle (1.41,0.); \node[text=white] at (-0.705,.15) {28\%}; \tikz \fill [niceRed2] (0.,0.3) rectangle (0.95,0.); \node[text=white] at (-0.475,.15) {19\%}; \end{tikzpicture}}  \\

         \arrayrulecolor{lightgray}
         \midrule
       TAS-Balanced &  472 & 0.97\% & .638 & \hspace{-.6cm} \adjustbox{valign=t}{\begin{tikzpicture}[baseline] \tikz \fill [niceGreen1] (0,0.3) rectangle (2.39,0.); \node[text=white] at (-1.195,.15) {48\%}; \tikz \fill [niceGreen2] (0.,0.3) rectangle (0.79,0.); \node[text=white] at (-0.395,.15) {16\%}; \tikz \fill [niceYellow2] (0.,0.3) rectangle (1.37,0.); \node[text=white] at (-0.685,.15) {27\%}; \tikz \fill [niceRed2] (0.,0.3) rectangle (0.44,0.); \node[text=white] at (-0.22,.15) {9\%}; \end{tikzpicture}} \\
       \arrayrulecolor{black}
       \bottomrule
    \end{tabular}
        \end{table}

\paragraph{\textbf{Study Design.}}
We select the queries from MSMARCO-DEV-49K with the largest RR@10 margin between BM25 and $\bertdot$.
For our analysis we set $\delta=1$ in Eq.~\ref{eq:robust-margin}, the maximum possible margin. Furthermore, to gain clarity on the actual result quality, without sparse label noise, we follow our labelling approach from Section \ref{sec:lexical-matches} to annotate the relevance of the first retrieved passage.

\paragraph{\textbf{Study Analysis.}}

In Table \ref{tab:bert_vs_bm25}, we show the results for standalone and TAS-Balanced trained dense retrieval models. BM25 shows a RR@1 and P@1 of 1 for the selected queries.
The number of queries where BM25 outperforms BERT$_\text{DOT}$ is 1.87\% using a standalone training and 0.97\% for TAS-Balanced. Judging from the sparse labels, this would already be a small number, albeit it could cause concern if the new search system fails on that many queries. However, our annotations show that while not all queries are answered, the true P@1 loss compared to BM25 for this query set is even smaller. The difference between standalone and TAS-Balanced shows that the small number of failures against BM25 are not endemic to dense retrieval, rather they can be even reduced further by better training techniques. 

If we set the threshold of allowed failures to a reasonable, albeit arbitrary, $1\%$ of queries we observe that TAS-Balanced passes our \cRobust tests (\appr) compared to BM25.

\subsection{Overall Decision}

When probing TAS-Balanced on the MS MARCO datasets, we surprisingly find that it passed all our robustness tests that were designed to target suspected flaws in dense retrieval. 
The main drawback of dense retrieval is the cost of vectorization and building the ANN index. In an application that needs to run at extremely low cost, the decision maker might put a higher weight on indexing size and time than they do on latency. They might also choose a point on the Pareto frontier, which loses significant NDCG@10 in order to greatly reduce cost. If 100 minutes of indexing and 22GB is impossible, for example on a very low-cost application, then the decision maker may choose the traditional index (tradeoff \ding{203}). However, we think that in many real applications the cost in minutes and gigabytes is affordable, so for our case study -- using MSMARCO -- we would expect the decision maker to mostly choose to replace term-based indexing with the new DR approach (tradeoff \ding{202}).

\section{Additional Criteria}
\label{sec:additional_criteria}
In addition to the criteria introduced in Section~\ref{sec:criteria}, there are many other important considerations, ranging from critical externalities to technical debt, that should be emphasized in any deployment decision.
In this section, we review several of these additional criteria that decision makers should pay attention to.
These considerations are also important in the context of dense and term-based retrieval systems and should be studied in future work.

\smallskip \noindent \cBias Search engines, like other large-scale information access systems, act as gatekeepers to the world's information.
The ranking of results that these systems produce directly influences what information and content users are exposed to.
When deploying new systems, it is therefore important to also consider potential representational and allocative harms that may result from biases encoded in these socio-technical systems.
For example, large deep learning models are known to not only pick up historical societal biases present in most training datasets, but can often amplify them, leading to harmful stereotyping of and promote negative sentiments towards marginalized groups~\citep{bender2021dangers}.
Biases in the models can also raise concerns around user-side and producer-side fairness~\citep{mehrotra2017auditing, ekstrand2021fairness}.
Therefore any system deployment decisions must ascertain that all benefits of the new system must be equitably distributed across different demographic groups and does not introduce any significant new societal harms.

\smallskip \noindent \cEnvironment As the worldwide scientific community rings the warning bell on critical dangers of continuing climate change~\citep{masson2021ipcc} and different institutions, commercial and otherwise, move towards more ambitious reduction of their carbon footprint and negative ecological impact (\eg, \citep{smith2020microsoft}), the environmental cost becomes an increasingly critical criteria to be considered in any deployment decisions.
The research community has recently raised various concerns on the environmental impact of large-scale deep learning models~\citep{bender2021dangers}. Same arguments also hold when these models are applied to retrieval models. Therefore, the environmental impact, \eg, carbon footprint, of the deployed system should be also considered in the decision making process. Many of these impacts are measurable, \eg, see \cite{Strubell2020}.

\smallskip \noindent \cMaintain Commercial search engines are large and complex systems often deployed over large-scale distributed cloud infrastructure.
In this paper, we have focused on the cost and benefit of a new system measured at the point of deployment, which misses the additional cost associated with maintaining and further improving the system post deployment.
For a machine-learned dense retrieval system, this includes the cost of incremental updates to the index as new documents are discovered and added to the collection, as well as old document contents are refreshed and re-indexed.
Over time, as the volume and the distribution of documents in the collection and the query workload naturally evolve, it may become necessary to periodically re-train and re-deploy the machine-learned models.
This introduces additional efficiency-effectiveness tradeoffs associated with future maintenance of any deployed system that decision makers must account for.
Neglecting these considerations can result in build up of technical debts in machine learning based systems~\cite{sculley2015hidden}.

\section{Conclusions}

Given the recent popularity of dense retrieval models, this paper takes a deeper look to evaluate these models and provide tools to answer the main question: \textit{Are we there yet? Can we switch from a term-based to a dense retrieval system?} We described a general framework for evaluating retrieval models based on multiple criteria that cover effectiveness, efficiency, and robustness measures.

In our case study, we observed that state-of-the-art ANN models can provide significantly higher effectiveness with similar query latency to term-based models. However, they have substantially higher cost in terms of storage usage and indexing time. Surprisingly, we observed that dense retrieval models deliver consistently better search results for queries with different length, different query term frequency, and robust lexical match capabilities.

More broadly for the field and in practical applications, the answer to the question \textit{are we there yet}, will be determined by many evaluations, many case studies. The proposed decision framework provides a guideline for systematic evaluation of new retrieval models which can be used by search engine practitioners to make informed decision about new retrieval models before deploying them into production. To say whether DR becomes a ubiquitous solution, perhaps even supplanting traditional indexing, requires many decisions to be made and many application-specific scenarios. We are certainly not there yet, but through many such studies we will find out.

\bibliographystyle{ACM-Reference-Format}
\bibliography{references}

\appendix

\section{Appendix: Impact of Cost Weighting}
\label{sec:appendix-weighting}

\begin{figure}[h]
       \includegraphics[trim={0cm 0cm 0cm 0cm},width=0.49\textwidth]{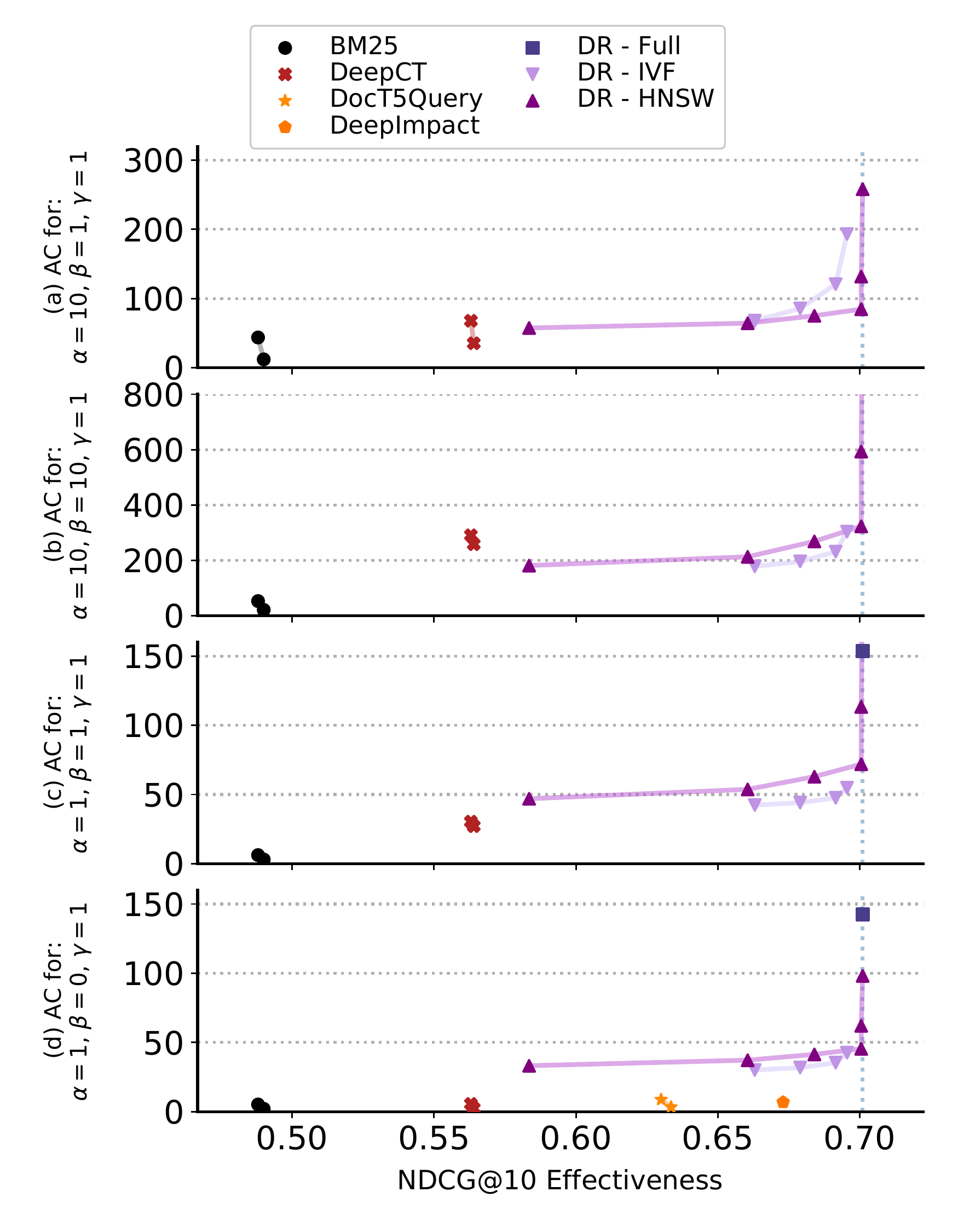}
    \centering
        \caption{Comparing different cost aggregation strategies on TREC-DL passage retrieval comparison}
    \label{fig:Pareto-variants}
    \end{figure}

In Section \ref{sec:primary-criteria}, we define a simple cost aggregation function (see Eq. \ref{eq:cost-aggregation}) with three parameters that control the weight of latency, indexing time, and storage requirement costs. In Figure \ref{fig:Pareto-variants}, we showcase additional weighting combinations in addition to the scenario in Figure \ref{fig:Pareto}. Many of the benchmarked methods have diametrical cost behaviors. While HNSW trades indexing time and storage space for lower latency, IVF does the opposite. The two methods based on T5 (DocT5Query \& DeepImpact) have enormous indexing times with very low latency and storage requirements.

We find that by modulating the weighting, a decision maker would come to different conclusions, as the combined Pareto frontier is set by different methods in Figure \ref{fig:Pareto-variants}. Note that these parameters are collection- and task- and domain-specific and should be carefully selected by domain experts and decision makers.

\begin{itemize}[leftmargin=*]
    \item \textbf{Figure \ref{fig:Pareto-variants} (a)} repeats the results of Figure \ref{fig:Pareto} (d) for easier comparison. It models an emphasis on query latency, and equally includes indexing time and storage.
    \item \textbf{Figure \ref{fig:Pareto-variants} (b)} increases the emphasis on indexing time. This is important for scenarios with many updates and index refreshes. We demonstrate that in this case the IVF method shows a better tradeoff than HNSW.
    \item \textbf{Figure \ref{fig:Pareto-variants} (c)} sets all weights to 1, which again favors IVF over HNSW and the difference between BM25 and the neural approaches becomes stronger, requiring a greater permissible cost increase factor for a selection of dense methods.
    \item \textbf{Figure \ref{fig:Pareto-variants} (d)} shows a scenario of a static collection, where indexing is a dismissible (and therefore ignored) one-time cost. The only cost factors that matter are storage space and latency. Now, the neural augmented term-based methods become much more viable to select, as their main drawback is the slow inference over all passages. 
\end{itemize}

This analysis again shows how we will not arrive at a single general recommendation for or against deploying a certain system. It depends strongly on the individual situation. Our framework provides the guidance for informed decision makers. 

\end{document}